\documentclass[10pt, a4paper]{article}
\usepackage{hyperref}
\usepackage{epstopdf}
\usepackage{stackrel}
\usepackage{graphicx}
\usepackage[utf8]{inputenc}
\usepackage{amssymb}
\usepackage{amsthm}
\usepackage{amsmath, calc}
\usepackage[authoryear]{natbib}

\begin{document}

\title{plsRglm: Partial least squares linear and generalized linear regression for processing incomplete datasets by cross-validation and bootstrap techniques with R.}
\author{Fr\'ed\'eric Bertrand\,$^{1,2,}$\footnote{to whom correspondence should be addressed} and Myriam Maumy-Bertrand\,$^{1,2}$\\\vspace{6pt} $^{1}$Labex IRMIA, Strasbourg, France.\\ $^{2}$IRMA - CNRS - UMR7501 - Universit\'e de Strasbourg\\ 67084 Strasbourg Cedex, France.}

\maketitle

\begin{abstract}

\subsubsection*{Summary:}
The aim of the \verb+plsRglm+ package is to deal with complete and incomplete datasets through several new techniques or, at least, some which were not yet implemented in \textbf{R}. Indeed, not only does it make available the extension of the PLS regression to the generalized linear regression models, but also bootstrap techniques, leave-one-out and repeated $k$-fold cross-validation. In addition, graphical displays help the user to assess the significance of the predictors when using bootstrap techniques. Biplots (Fig. \ref{fig:SI1}) can be used to delve into the relationship between individuals and variables.

\subsubsection*{Availability:}
\href{http://cran.r-project.org/web/packages/plsRglm/index.html}{{\tt plsRglm}} is freely available from the \textbf{R} archive CRAN. The package is distributed under the GNU General Public License (version 3 or later) and includes vignettes, example files and datasets.

\subsubsection*{Contact:} \href{fbertran@math.unistra.fr}{fbertran@math.unistra.fr}
\subsubsection*{Supplementary information:}\href{http://cran.r-project.org/web/packages/plsRglm/vignettes/plsRglm.pdf}{Vignette} and \href{http://cran.r-project.org/web/packages/plsRglm/vignettes/plsRglm-manual.pdf}{manual} of the package.
\end{abstract}

\section{Motivation}
Extracting knowledge from datasets featuring a large number of variables is more and more a common task, especially in medicine or in biology. However, some known issues, for instance, a strong collinearity between the predictors not to mention landscape datasets -more predictors than subjects- or missing data, usually lead us to use other statistical estimation methods than the usual least squares to fit a linear regression model. One of these methods is the Partial Least Square (PLS) Regression (PLSR) which was first introduced by \cite{wold1983multivariate} and \cite{Wold:1984}. This estimation method was already implemented in several R 
or Bioconductor packages. 
Despite recent progresses on degrees of freedom correction (DOFc) made by \cite{krsu2011} (R package \verb+plsdof+), all the other packages only provide cross-validation based criteria for selecting the number of PLSR components. Moreover, all the packages lack missing data support, either issuing an error or abruptly removing rows with NA values. They also lack bootstrap techniques even for complete datasets.

As a consequence, the key features of the \verb+plsRglm+ package are to deal either with complete or incomplete datasets, to support both PLSR -regular or weighted- and its extension to generalized linear regression models (PLSGLR) by \cite{bastien2005pls} and provide bootstrap techniques \citep{lazraq2003selecting,bastien2005pls} to assess the significance of the original predictors -graphical outputs were designed to help to proceed this step- in all of these settings. In addition, various criteria, either cross-validation based (including $PRESS$, $Q^2$, $Q^2$cum, $PRE\chi2$, $\chi^2$, $Q^2\chi^2$ or misclassified) or not ($AIC$, $BIC$, significance tests, all three with DOFc when available), make it easier to pick out a relevant number of components, thus filling several gaps and providing R users with state-of-the-art tools to process real complex datasets. 

We deal in the vignette with 7 example datasets, including a study on the quality of some Bordeaux wines \citep{Tenenhaus} (ordinal logistic PLSR), the Cornell dataset \citep{kettaneh1992analysis} (regular PLSR) or an allelotyping dataset on which we applied a binomial logistic PLS regression model with success in \cite{meyer2010comparaison}.

\section{Application}
\subsection{A binary incomplete dataset}

To illustrate the package features, we processed a colon cancer allelotyping study \citep{meyer2010comparaison}. This data set, called \verb+aze+, was collected on patients suffering from a colon adenocarcinoma. It features 104 observations of 33 binary qualitative explanatory variables (the microsatellites), and one binary response, $y$ (a binary cancer stage derived from the Astler-Coller classification \citet{asco54}). Due to unavoidable technical limitations, about one third of the data are missing. 

\subsection{Model selection and cross-validation}

 The response being a two-level factor, we decided to apply a binary logistic PLS regression model with a $logit$ link-function. First step, we used cross validation to choose a relevant number of components for the PLS model: a 100 repeated $8$-fold cross validation (random groups of 13 observations) was applied to the data set. For the first split, results are displayed in Table~\ref{table:cv8}.
\begin{table}[!t]
{\begin{tabular}{lcccc}\hline
  Nb components & 0 & 1 & 2 & 3\\\hline
  AIC & 145.83 & 119.06 & 105.96 & 100.28 \\
  BIC & 148.47 & 124.35 & 113.89 & 110.86 \\
  Miss Classed & 49 & 30 & 20 & 18 \\
  Significant pred. &  & 1 & \textbf{3} & 0 \\
  Miss Classed (8-CV) &  & 58 & 62 & 56 \\
  $Q^{2}\chi^{2}$ (8-CV) &  & -5.29 & -8.54 & -34.87 \\
  $\chi^{2}$ Pearson & 104.00 & \textbf{101.7}1 & 110.98 & 102.52 \\\hline
\\
	Nb components & 4 & 5 & 6 & 7\\\hline
  AIC & 96.2 & 94.17 & \textbf{93} & 94.11 \\
  BIC & \textbf{109.42} & 110.04 & 111.51 & 115.26 \\
  Miss Classed & 20 & 18 & \textbf{16} & 17\\
  Significant pred. & 0 & 0 & 0 & 0\\
  Miss Classed (8-CV) & \textbf{55} & 56 & 63 & 64\\
  $Q^{2}\chi^{2}$ (8-CV) & $-2.38*10^{2}$ & $-2.43*10^{3}$ & $-1.41*10^{6}$ & $-9.26*10^{9}$\\
  $\chi^{2}$ Pearson & 122.84 & 148.72 & 141.1 & 149.1\\\hline
\end{tabular}}{}
\caption{Results of cross-validation, k=8}
\label{table:cv8}
\end{table}

\begin{figure}[!tp]
    \centering
    \includegraphics[trim = 0cm 0cm 0cm 0cm, clip,scale=0.75]{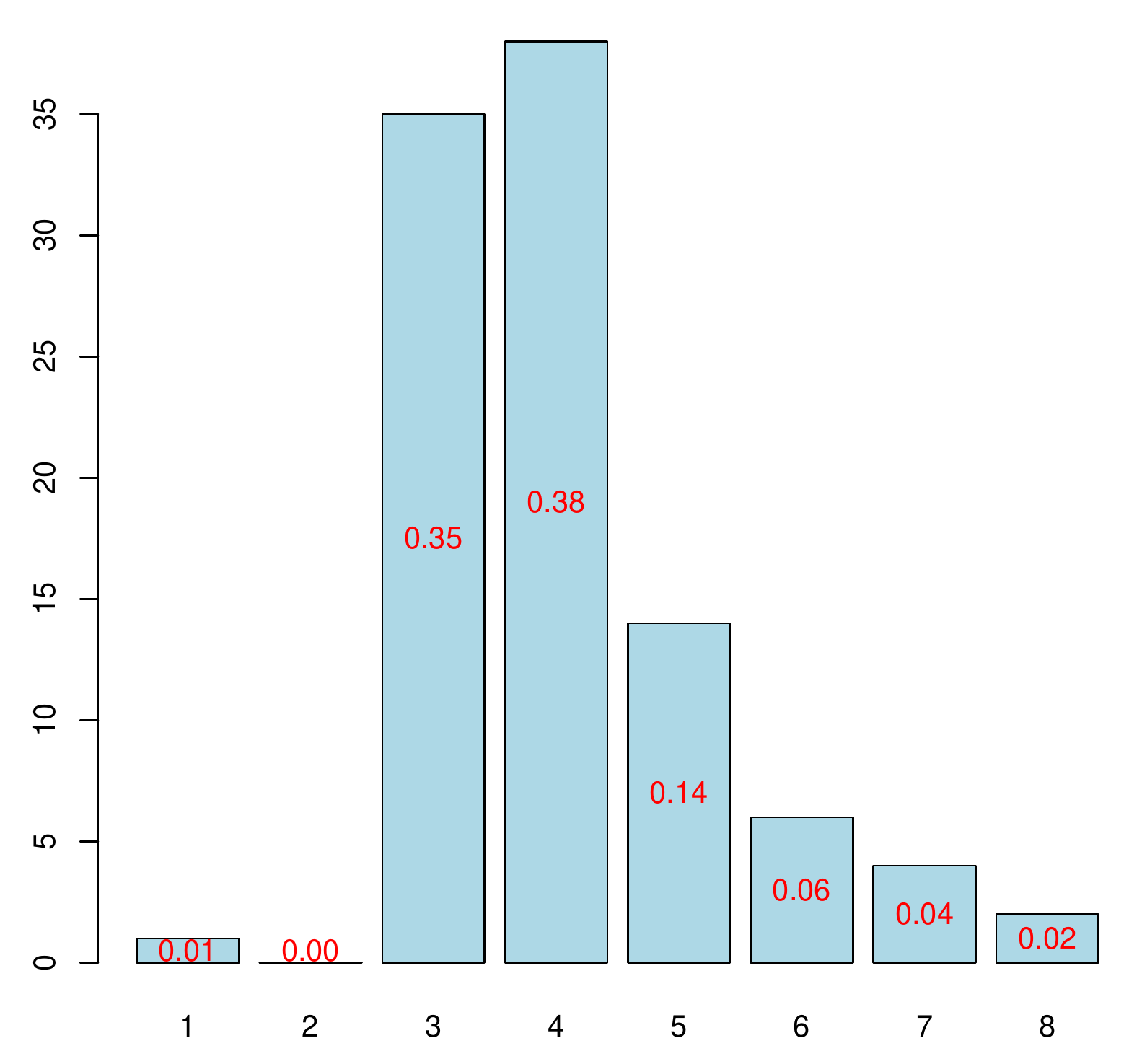}
		\caption{ Nb components, 8-CV, n=100}\label{fig:1a}
    \end{figure}

According to the number of significant predictors within each component (Significant pred., \citet{bastien2005pls}), we should retain 2 components, whereas using the BIC this number raises to 4 and using the AIC to 6. For this study, we decided to use the cross validated number of misclassified observations criteria. The complete results of the 100 repeated cross-validation are displayed on Figure \ref{fig:1a}: it suggests to retain 4 components, in agreement with the BIC criteria. Denoting by $t_h$ the $h^{th}$ PLS component, $c_h$ its coefficient and $\mu$ the intercept, the PLS binary logistic model is:
\begin{equation}
       \mathbb{P}\left(y=1\right)=\left({e^{\mu+\sum_{h=1}^4c_ht_h}}\right)\Big/\left({1+e^{\mu+\sum_{h=1}^4c_ht_h}}\right).
\label{mod1}
\end{equation}
The estimates of the coefficients of this PLSGLR are $\hat c_1=1.4274$, $\hat c_2=0.5096$, $\hat c_3=0.6903$, $\hat c_4=0.7930$ and $\hat\mu=-0.2968$. One can then convert these estimates to those of the coefficients of the original predictors.


\subsection{Significance of predictors and bootstrap}
\begin{figure}[!tp]
    \centering
    \includegraphics[trim = 0cm 0cm 0cm 0cm, clip,scale=0.75]{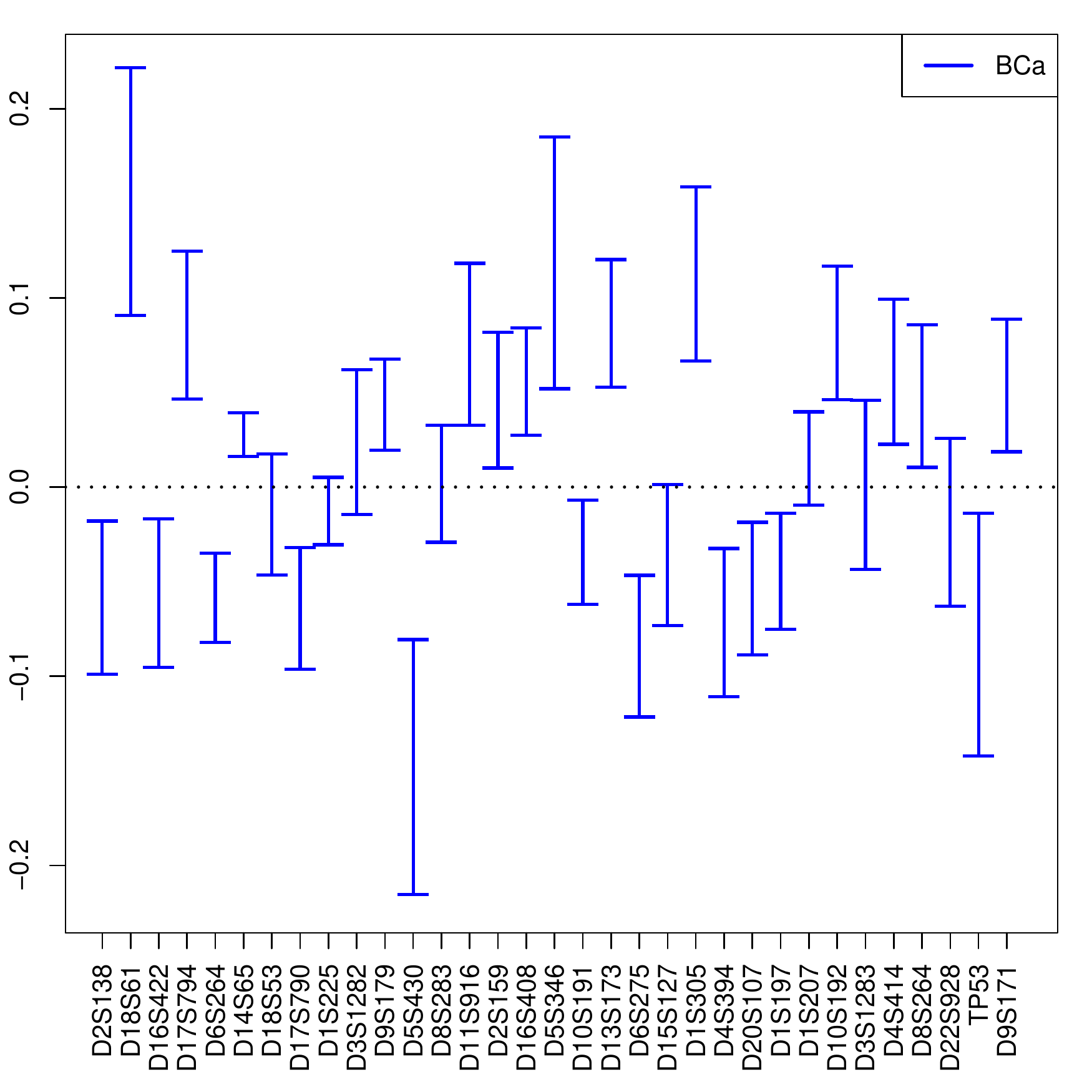}
		\caption{ $BC_a$ Bootstrap CI of regressor's coefficients.}\label{fig:1b}
    \end{figure}

\begin{figure}[!tp]
    \centering
    \includegraphics[trim = 0cm 3.4cm 0cm 0cm, clip,scale=0.75]{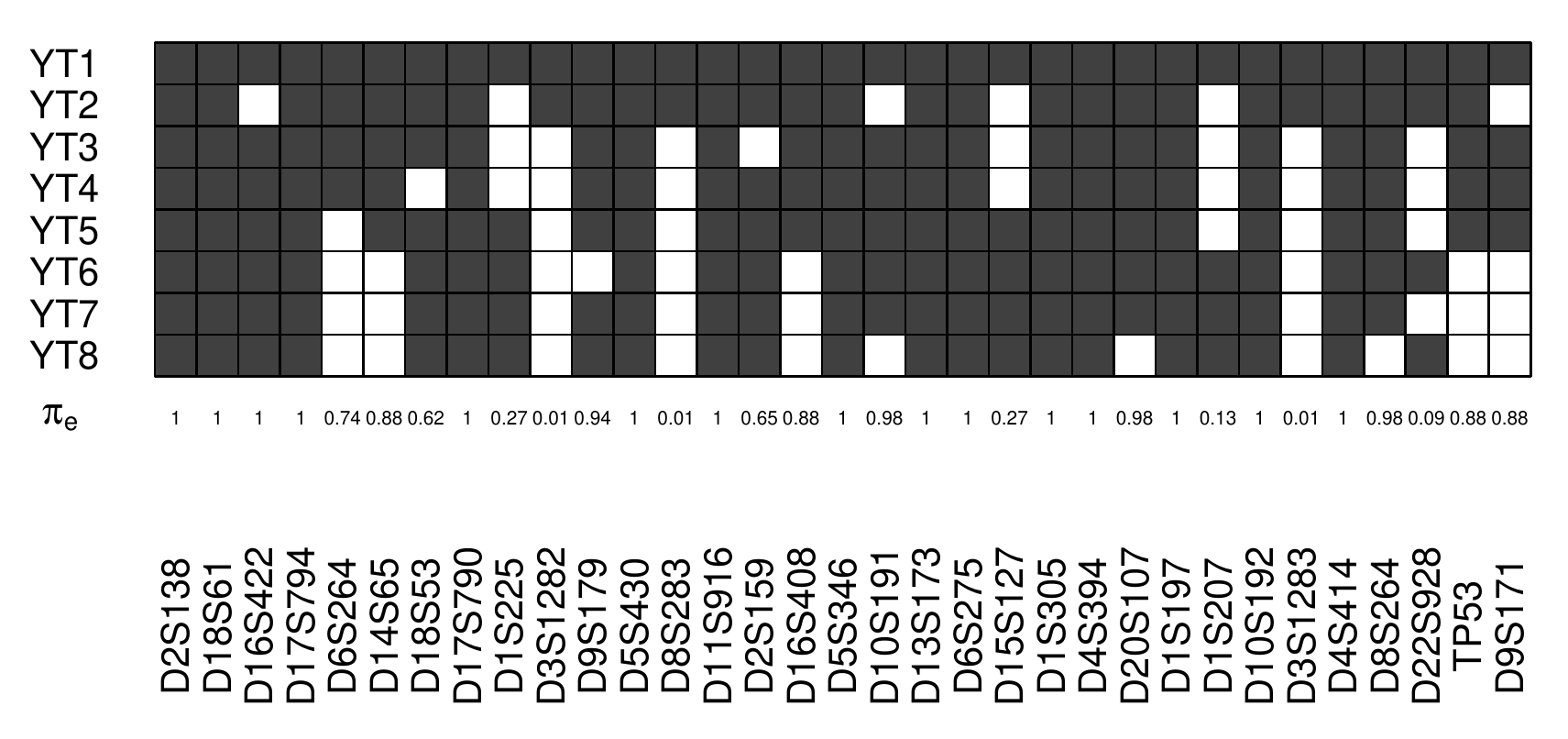}
		\caption{ Significant predictors through the models with 1, 2, 3, 4, 5 and 6 components and robust significance index $\pi_e$.}\label{fig:1c}
    \end{figure}

\begin{figure}[!tp]
\includegraphics[trim = 0cm 0cm 0cm 0cm, clip,scale=.75]{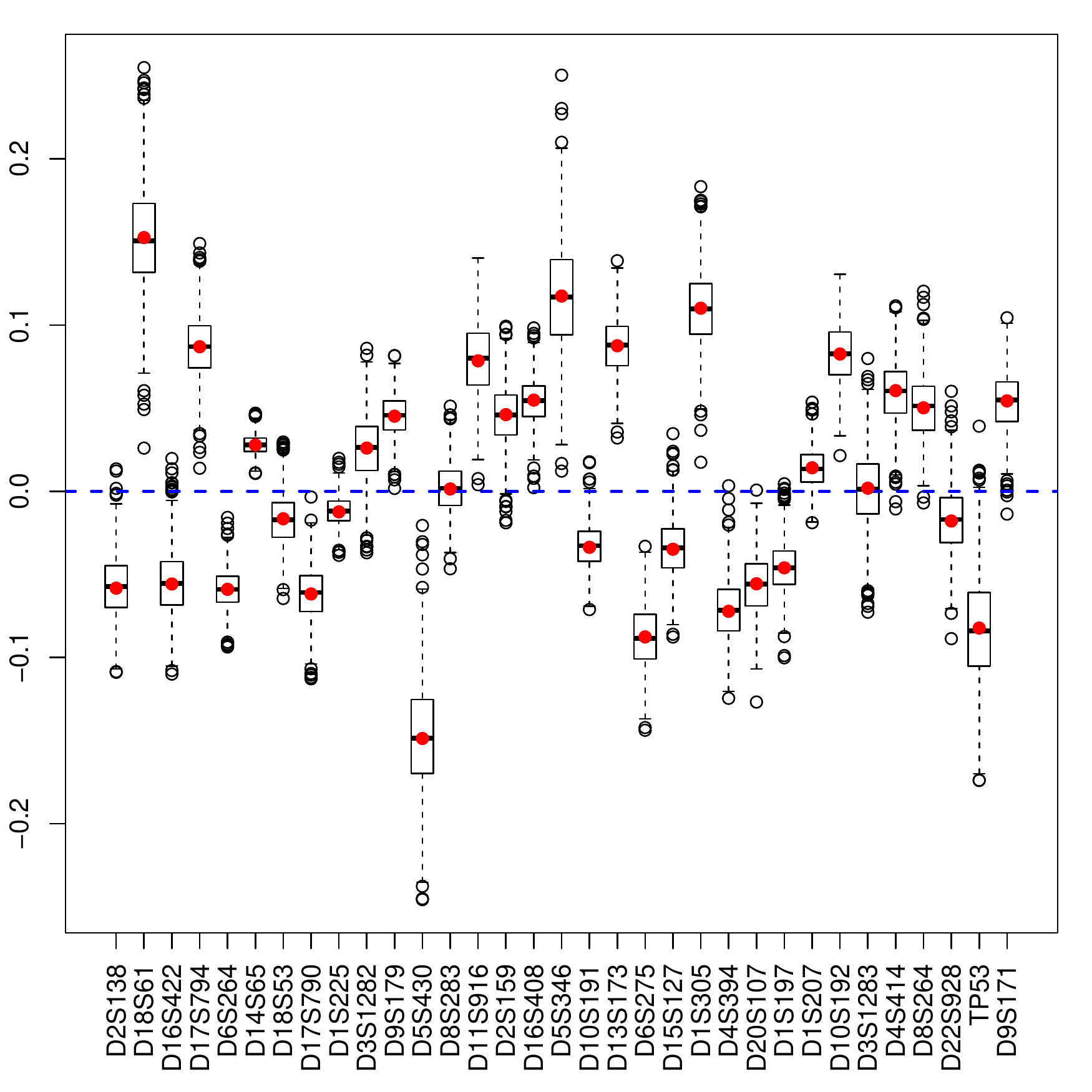}
\caption{Boxplots of the $(Y,T)$ bootstrap distribution of the predictors (4 components model and 1000 resamplings).}\label{fig:SI2}
\end{figure}

\begin{figure}[!tp]
\includegraphics[trim = 0cm 0cm 0cm 0cm, clip,scale=.75]{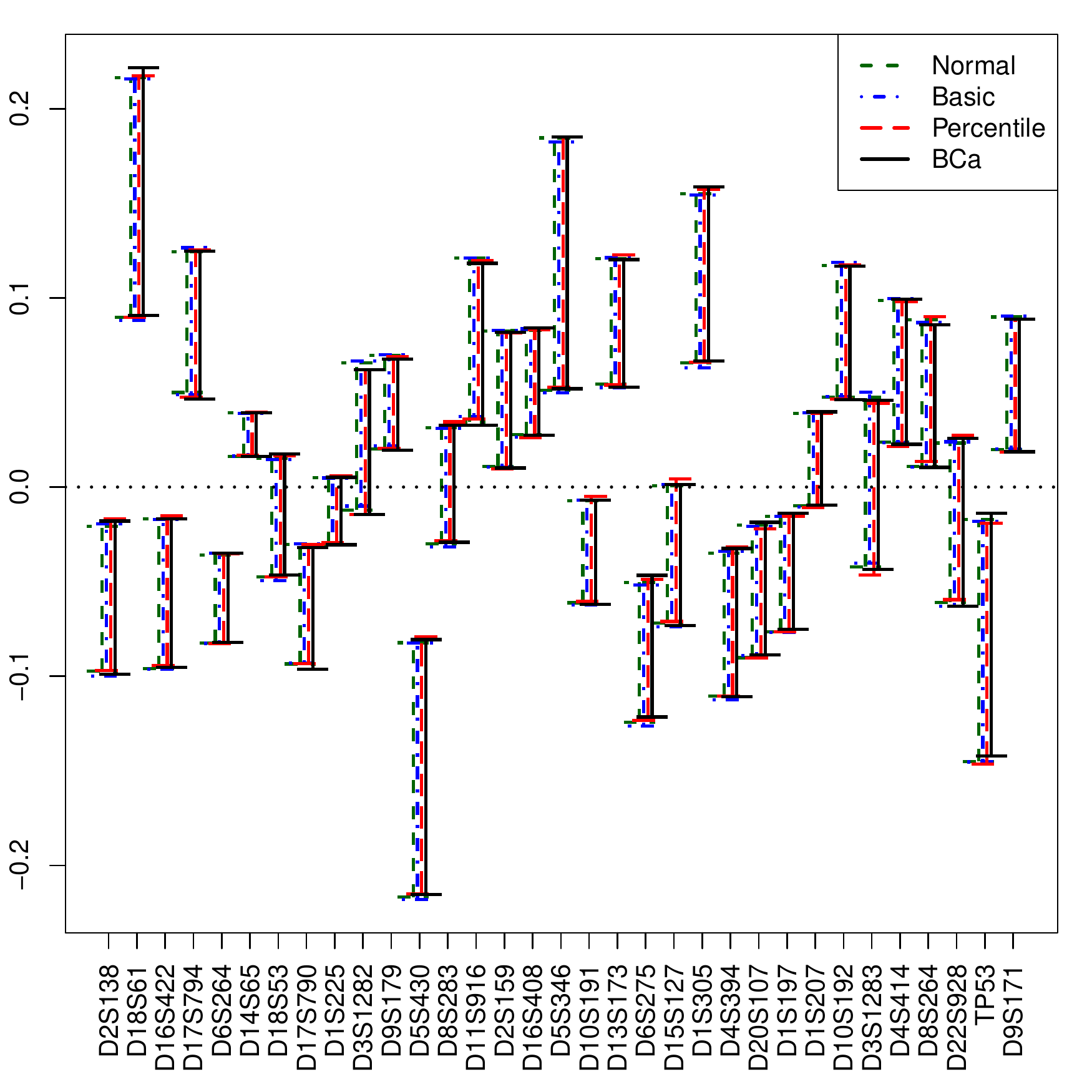}
\caption{Four types of $(Y,T)$ Bootstrap CI of regressor's coefficients (4 components model and 1000 resamplings).}\label{fig:SI3}
\end{figure}

\begin{figure}[!tp]
\includegraphics[trim = 0cm 0cm 0cm 0cm, clip,scale=.75]{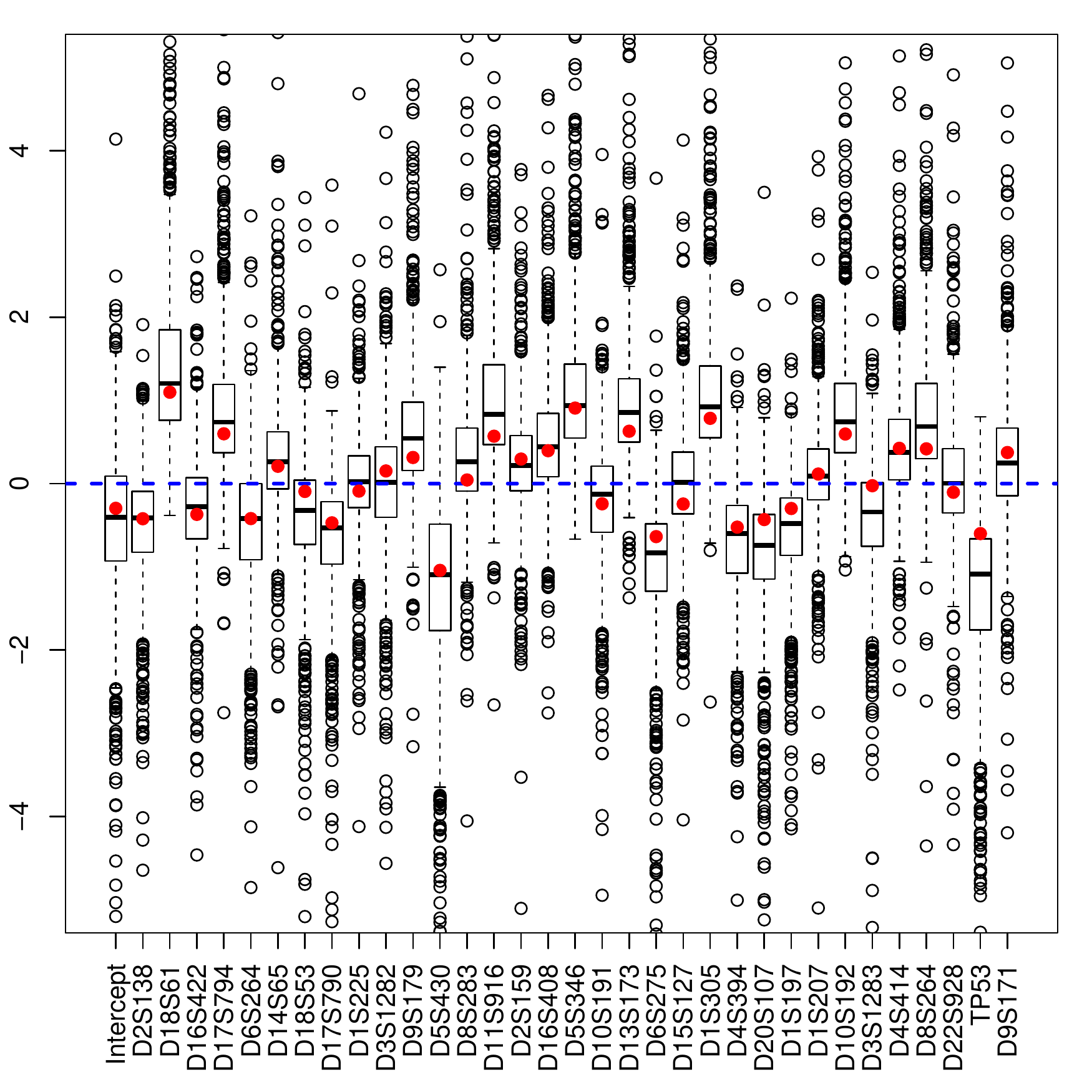}
\caption{Boxplots of the $(Y,X)$ bootstrap distribution of the predictors (4 components model and 1000 resamplings).}\label{fig:SI4}
\end{figure}

\begin{figure}[!tp]
\includegraphics[trim = 0cm 0cm 0cm 0cm, clip,scale=.75]{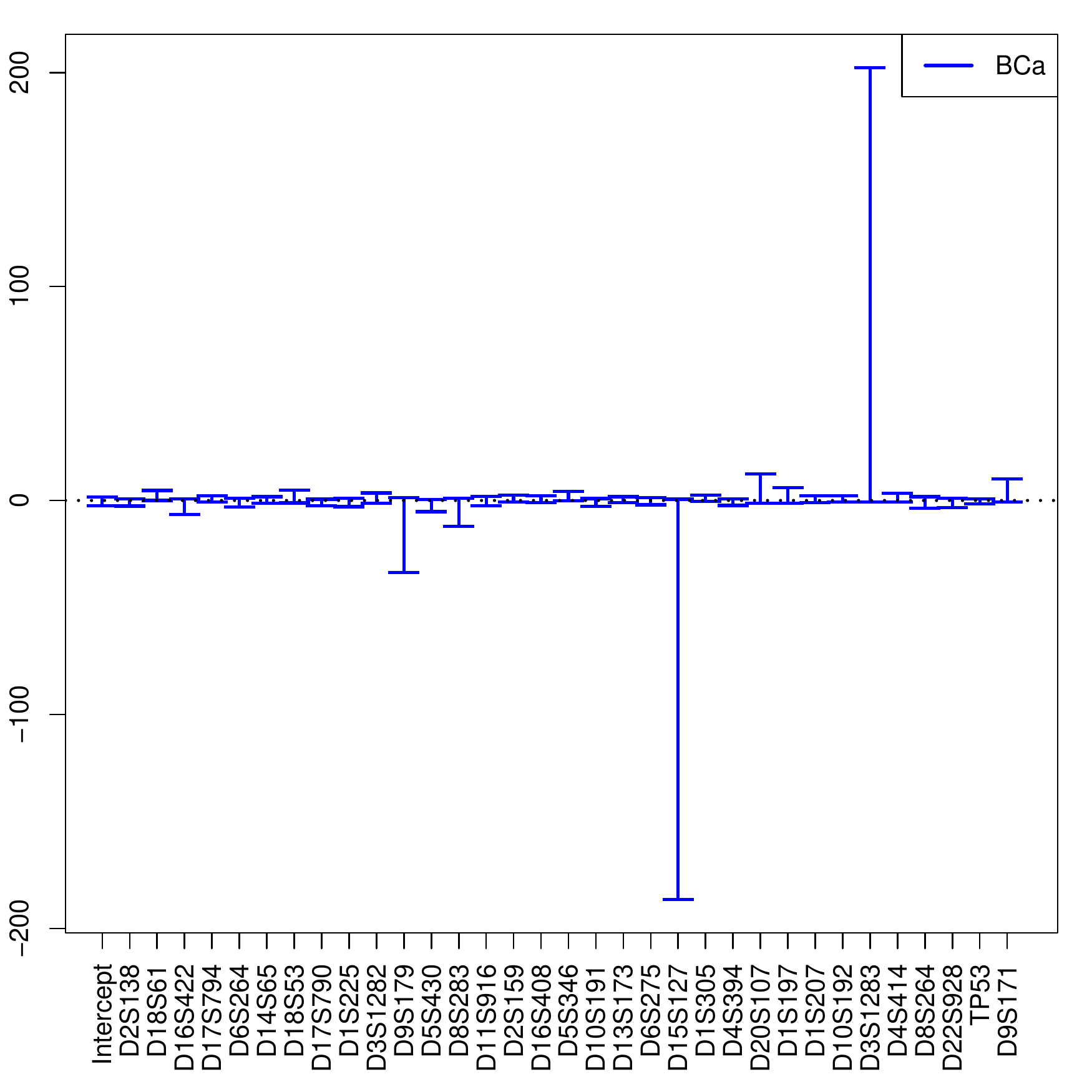}
\caption{$BC_a$ $(Y,X)$ Bootstrap CI of regressor's coefficients (4 components model and 1000 resamplings).}\label{fig:SI5}
\end{figure}

The next issue to be tackled is to pick significant predictors $\mathbf{x}_j,\; 1\leqslant j\leqslant 33$ and thus estimate the coefficients $\beta_j$ of the predictors using the four component model (see Figs \ref{fig:SI2} to \ref{fig:SI5}).  Two resampling schemes for bootstrap are implemented in the package: $(Y,X)$ \citep{lazraq2003selecting} or $(Y,T)$  \citep{bastien2005pls}. For this study, we chose the latter and performed 1000 resamplings: boxplots of the bootstrap distribution can easily be plotted using the package (Fig. \ref{fig:SI2}) as well as $BC_a$ confidence intervals (CI) for each of the predictors (Fig. \ref{fig:1b}). We focused on $BC_a$ CI, since these are usually recommended and even though the package can derive CI with percentile, normal or basic bootstrap (Fig. \ref{fig:SI3}). Only 9 predictors did not differ significantly from 0 at the 5\% level. It is worth noting, as reported on Figure \ref{fig:1c}, that some of the predictors feature a stability property as they are significant in all the models with 1 to 8 components for $BC_a$ CI based on 1000 $(Y,T)$ resamplings.

Moreover, as to the significance of predictors, there are few differences between the models with 3 and 4 components as only 1 predictor, significant in former model, becomes non-significant in the latter model, the converse being also true. This should be stressed if one recalls cross-validation results: the 4 components model (40\%) was only slightly ahead of the 3 components one (35\%). Computing the empirically weighted, with respect to the CV distribution of components, proportion of the models for which the predictors are significant, yields a robust significance index $\pi_e$ (reported on Figure \ref{fig:1c}) with respect to a number of components misspecification.

One must know that, even though the $(Y,T)$ bootstrap technique is clearly faster and more stable than the $(Y,X)$ one, especially in a GLR context, these two techniques lead to dramatically different results (Fig. \ref{fig:SI4} and \ref{fig:SI5}).





\section{Conclusion}

The scope of application of the \verb+plsRglm+ package is steadily widening as the complexity of the datasets has been  increasing for several years. Among many others, medicine, biology and chemistry are domains where it is likely to have to deal with issues like strongly correlated predictors or even rectangular datasets featuring more predictor than observations in a linear or a generalized linear regression. For instance, technological breakthroughs, such as next-gen sequencing, favors the use of generalized linear regression models such as Poisson, quasi-Poisson or negative binomial ones, with rectangular datasets. PLS extensions of all these three models are available in the \verb+plsRglm+ package. Similar issues are unavoidable with mixture modeling, spectrum analysis or other omics data analysis.

In a word, we view this package as a key additional toolbox for the \textbf{R} language.


\bibliographystyle{plainnat}
\bibliography{biblio}

\begin{figure}[!tp]
\includegraphics[trim = 0cm 0cm 0cm 0cm, clip,scale=.75]{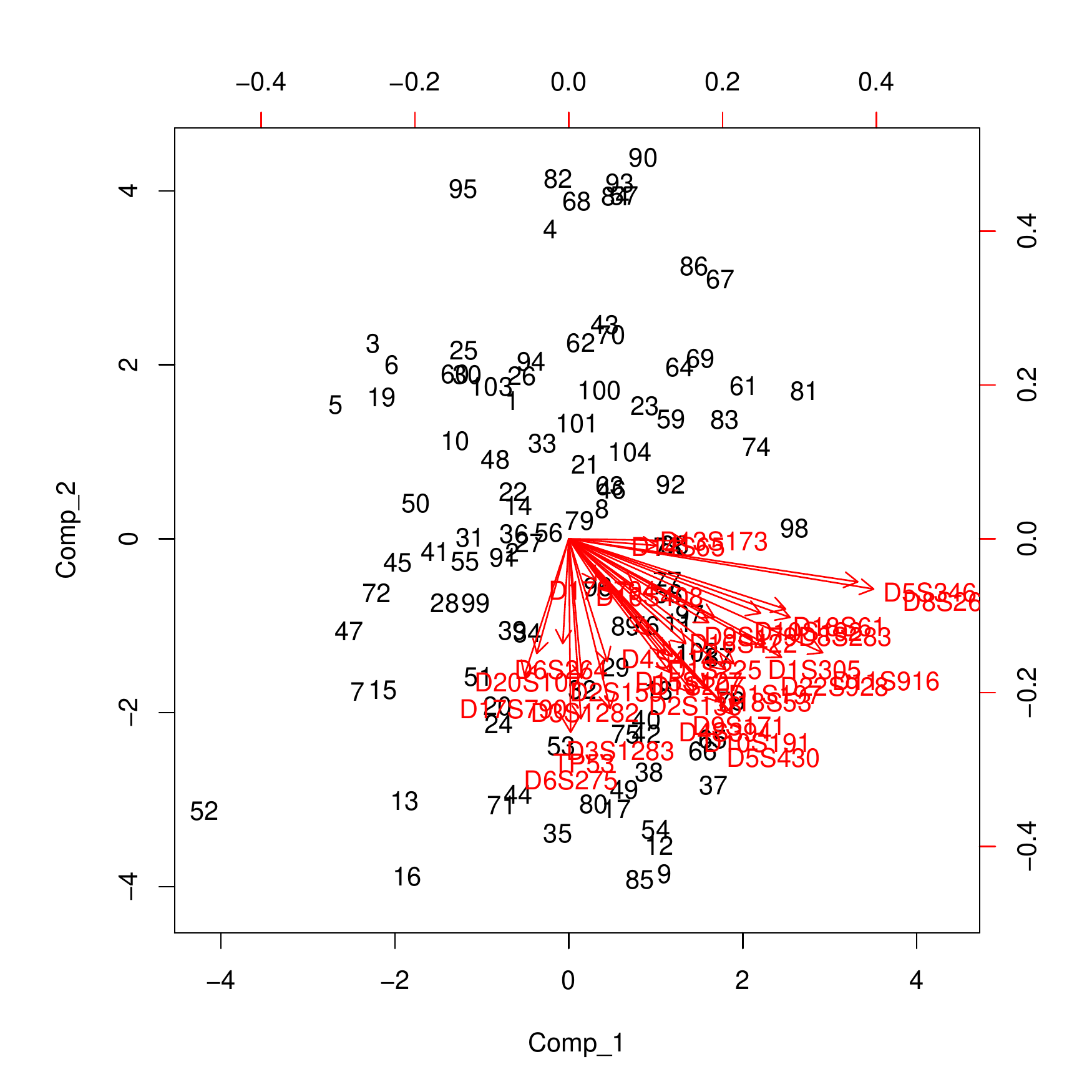}
\caption{Biplot of individuals and variables (first two components).}\label{fig:SI1}
\end{figure}

\end{document}